\documentclass[preprint2]{aastex}

\slugcomment{Accepted for publication in the Astrophysical
 Journal (Letters) on 5 October 2005.}

\newcommand{\etal}{et~al.}

\newcommand{\Msun}{M_{\odot}}
\newcommand{\kms}{km~s$^{-1}$}
\newcommand{\ergs}{erg~s$^{-1}$}
\newcommand{\Fefs}{$^{56}$Fe}
\newcommand{\Cofs}{$^{56}$Co}
\newcommand{\Nifs}{$^{56}$Ni}
\newcommand{\Mms}{M_{\rm MS}}
\newcommand{\Mej}{M_{\rm ej}}

\def\gsim{\mathrel{\rlap{\lower 4pt \hbox{\hskip 1pt $\sim$}}\raise 1pt
\hbox {$>$}}}
\def\lsim{\mathrel{\rlap{\lower 4pt \hbox{\hskip 1pt $\sim$}}\raise 1pt
\hbox {$<$}}}

\begin{document}

\title{The Unique Type Ib Supernova 2005bf: A WN Star Explosion Model 
 for Peculiar Light Curves and Spectra}

\author{
 N.~Tominaga\altaffilmark{1},
 M.~Tanaka\altaffilmark{1},
 K.~Nomoto\altaffilmark{1,2},
 P.A.~Mazzali\altaffilmark{1,2,3,4},
 J.~Deng\altaffilmark{5},
 K.~Maeda\altaffilmark{6},
 H.~Umeda\altaffilmark{1},
 M.~Modjaz\altaffilmark{7},
 M.~Hicken\altaffilmark{7},
 P.~Challis\altaffilmark{7},
 R.P.~Kirshner\altaffilmark{7},
 W.M.~Wood-Vasey\altaffilmark{7},
 C.H.~Blake\altaffilmark{7},
 J.S.~Bloom\altaffilmark{8},
 M.F.~Skrutskie\altaffilmark{9},
 A.~Szentgyorgyi\altaffilmark{7},
 E.E.~Falco\altaffilmark{7},
 N.~Inada\altaffilmark{10},
 T.~Minezaki\altaffilmark{10},
 Y.~Yoshii\altaffilmark{10},
 K.~Kawabata\altaffilmark{11},
 M.~Iye\altaffilmark{12},
 G.C.~Anupama\altaffilmark{13},
 D.K.~Sahu\altaffilmark{13},
 and T.P.~Prabhu\altaffilmark{13} 
 }

\altaffiltext{1}{Department of Astronomy, School of Science,
University of Tokyo, Bunkyo-ku, Tokyo 113-0033, Japan;
tominaga@astron.s.u-tokyo.ac.jp, mtanaka@astron.s.u-tokyo.ac.jp, 
nomoto@astron.s.u-tokyo.ac.jp, umeda@astron.s.u-tokyo.ac.jp}
\altaffiltext{2}{Research Center for the Early Universe, School of
Science, University of Tokyo, Bunkyo-ku, Tokyo 113-0033, Japan}
\altaffiltext{3}{Max-Planck Institut f\"ur Astrophysik,
Karl-Schwarzschild Str. 1, D-85748 Garching, Germany;
mazzali@MPA-Garching.MPG.DE}
\altaffiltext{4}{Istituto Nazionale di Astrofisica-OATs, Via Tiepolo 11,
I-34131 Trieste, Italy}
\altaffiltext{5}{National Astronomical Observatories, CAS, 20A Datun Road,
Chaoyang District, Beijing 100012, China; jsdeng@bao.ac.cn}
\altaffiltext{6}{Department of Earth Science and Astronomy,
Graduate School of Arts and Science, University of Tokyo, Meguro-ku, Tokyo
153-8902, Japan; maeda@esa.c.u-tokyo.ac.jp}
\altaffiltext{7}{Harvard-Smithsonian Center for Astrophysics, 60
Garden Street, Cambridge, MA, 02138, USA; mmodjaz@cfa.harvard.edu,
mhicken@cfa.harvard.edu, pchallis@cfa.harvard.edu,
kirshner@cfa.harvard.edu, wmwood-vasey@cfa.harvard.edu,
cblake@cfa.harvard.edu, saint@cfa.harvard.edu, efalco@cfa.harvard.edu}
\altaffiltext{8}{Department of Astronomy, University of California
Berkeley, 601 Campbell Hall, Berkeley, CA 94720, USA;
jbloom@astro.berkeley.edu}
\altaffiltext{9}{Department of Astronomy, University of Virginia,
Charlottesville, VA 22903, USA; mfs4n@virginia.edu}
\altaffiltext{10}{Institute of Astronomy, School of Science, University
of Tokyo, 2-21-1 Osawa, Mitaka, Tokyo 181-0015, Japan;
inada@ioa.s.u-tokyo.ac.jp, minezaki@mtk.ioa.s.u-tokyo.ac.jp, yoshii@mtk.ioa.s.u-tokyo.ac.jp}
\altaffiltext{11}{Hiroshima Astrophysical Science Center, Hiroshima
University, Hiroshima 739-8526, Japan; kawabtkj@hiroshima-u.ac.jp}
\altaffiltext{12}{Optical and Infrared Astronomy Division, National
Astronomical Observatory of Japan, 2-21-1 Osawa, Mitaka, Tokyo 181-8588,
Japan; iye@optik.mtk.nao.ac.jp}
\altaffiltext{13}{Indian Institute of Astrophysics, Koramangala,
Bangalore 560 034, India; gca@iiap.res.in, dks@crest.ernet.in, tpp@iiap.res.in}

\begin{abstract}

Observations and modeling for the light curve (LC) and  spectra of supernova
(SN) 2005bf are reported. This SN showed unique features: the LC had two
maxima, and declined  rapidly after the second maximum, while the spectra
showed  strengthening He lines whose velocity increased with time. The
double-peaked LC can be reproduced by a double-peaked \Nifs\ distribution, with
most \Nifs\ at low velocity and a small amount at high velocity. The rapid
post-maximum decline requires a large fraction of the $\gamma$-rays to escape
from the \Nifs-dominated region, possibly because of low-density ``holes''. The
presence of Balmer lines in the spectrum suggests that the He layer of the
progenitor was substantially intact.  Increasing $\gamma$-ray deposition in the
He layer due to enhanced $\gamma$-ray escape from the \Nifs-dominated region
may explain both the delayed strengthening and the increasing velocity of the
He  lines. The SN has massive ejecta ($\sim6-7\Msun$), normal kinetic energy
($\sim 1.0-1.5\times 10^{51}$  ergs), high peak bolometric luminosity ($\sim 
5\times 10^{42}$ erg s$^{-1}$) for an epoch as late as $\sim$ 40 days, and a
large \Nifs\ mass ($\sim0.32\Msun$). These properties, and the presence of a
small amount of H suggest that the progenitor was initially massive (M$\sim
25-30 \Msun$) and had lost most of its H envelope, and was possibly a WN star. The
double-peaked \Nifs\ distribution suggests that the explosion may have formed
jets that did not reach the He layer.  The properties of SN 2005bf resemble
those of the explosion of Cassiopeia A.

\end{abstract}


\section{INTRODUCTION}
\label{sec:intro}

As the sample of well-studied supernovae (SN) grows, it is not surprising that
rare types are observed.  At first, these may seem strange and unique; in time
they may become well-understood subtypes.  We believe that SN 2005bf is one of
these new types: it has unique photometric and spectroscopic behavior.  Based
on a successful effort to model our light curve and spectra, we believe that SN
2005bf fits into the scheme suggested by Nomoto et al. (1995) which places
core-collapse SN in a sequence (IIP-IIL-IIb-Ib-Ic) of increasing mass loss from
the progenitor star.

\begin{figure*}
\epsscale{1.8}
\plotone{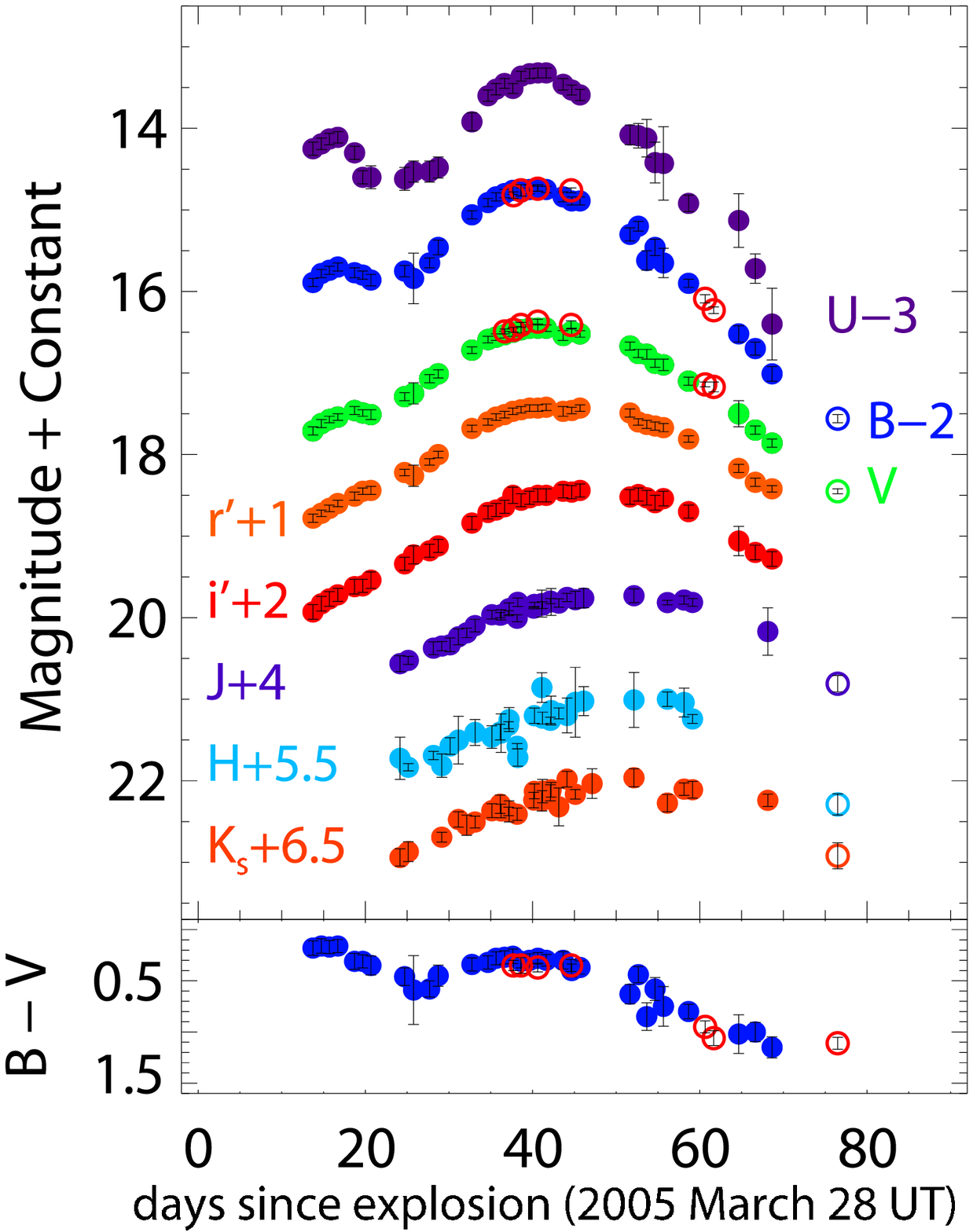} 
\caption{(top) Observed $UBVr'i'JHK_s$ (or $K$) light curve of SN~2005bf. Filled symbols are 
data from the Fred Lawrence Whipple Observatory (\citealt{mod05c}), open 
symbols are data from the Himalayan Chandra Telescope and the MAGNUM Telescope 
(\citealt{anu05,ina05}). 
(bottom) Observed $B-V$ color evolution. The meaning of the symbols is the 
same as in the top panel. \label{fig:obsLC}}
\end{figure*}

Discovered by Monard (2005) and Moore \& Li (2005) on April 6, 2005 (UT) in the
spiral arms of the SBb galaxy MCG +00-27-5, SN 2005bf was initially classified
as a Type Ic SN (SN Ic) (Morell et al. 2005; Modjaz et al. 2005a).  The later development of He lines
suggested an unprecedented transition to Type Ib (Wang \& Baade 2005; Modjaz et
al. 2005b). Even stranger, the light curve was very different from any known SN
(Hamuy et al. 2005): a fairly rapid rise to a first peak was followed by a
period of stalling or slow decline and by a new rise to a later, brighter peak
at $\sim 40$ days after explosion (Fig.~\ref{fig:obsLC}).  The brightness
($M_{\rm bol}\sim -18$ mag) and the late epoch of the second peak suggest this
SN ejected a large amount of \Nifs. SN 2005bf does not show the broad lines
seen in hypernovae.  These properties make SN~2005bf a very interesting SN.  

Photometry was obtained in $UBVr'i'$ bands at the 1.2m telescope at the Fred
Lawrence Whipple Observatory on Mt. Hopkins (FLWO; Modjaz et al. 2005c),
$JHK_s$ bands at the refurbished 1.3m telescope on Mt. Hopkins (PAIRITEL;
http://pairitel.org; Modjaz et al. 2005c) at FLWO, $BVRI$ bands at the
Himalayan Chandra Telescope (HCT; \citealt{anu05}), and $BVRIJHK$ bands at the
MAGNUM Telescope \citep{yos03,ina05} (Fig.~\ref{fig:obsLC}).  Spectroscopic
observations were made at FLWO, HCT and the SUBARU Telescope
\citep{kaw02,kaw05}. We used the light curve and spectra to constrain models
for the SN and to infer the properties of the progenitor star\footnote[14]{After this paper was submitted and placed on astro-ph, the Carnegie 
Supernova Project and their collaborators placed a paper on SN 2005bf 
on astro-ph \citep{fol05}.  It is beyond the scope of this Letter to 
compare the observations and theoretical models in detail.}.

\section{LIGHT CURVE MODELS}
\label{sec:LC}

The bolometric light curve (LC) was constructed as in Yoshii et
al. (2003; see Fig.~\ref{fig:LC}). Synthetic bolometric LCs were computed with an LTE
radiation hydrodynamics code and a gray $\gamma$-ray transfer code
\citep{iwa00}. Electron-scattering and line opacity were considered for
radiation transport. The former used the ionization computed with the Saha
equation. For the latter the approximate formula given in \cite{maz01} was
adopted, with modifications to account for the C, O, and He-rich layers. We
assumed a Galactic reddening $E(B-V)=0.045$, a distance modulus $\mu=34.5$
(\citealt{sch98};  http://nedwww.ipac.caltech.edu/index.html; \citealt{fal99}),
and an explosion date of $2005$ March 28 UT as inferred from the marginal
detection on 2005 March 30 UT \citep{moo05}.  

The LC is powered by the radioactive decay of \Nifs\ to \Cofs\ and \Fefs.  The
theoretical LC width near peak depends on ejected mass $\Mej$ and  explosion
kinetic energy $E$ as $\Mej E^{-3}$ \citep{arn82,nom04}.  The mass and
distribution of \Nifs\ are constrained by the LC brightness and shape, but
various combinations of ($\Mej$, $E$) can fit the LC.  Spectra break this
degeneracy and help establish the abundance distribution in the ejecta.

The density structure used for the LC calculation was based on the C+O star
model CO138E50 used for SN~1998bw (\citealt{nak01}) but rescaled by changing
$r$, $v$, and $\rho$ such that $\Mej$ and $E$  are rescaled as $\Mej \propto
\rho r^3$ and $E \propto \rho v^2 r^3$, respectively. The He distribution is
constrained by observational evidence and spectral fitting, as discussed in \S
3.

\begin{figure*}
\epsscale{2.}
\plotone{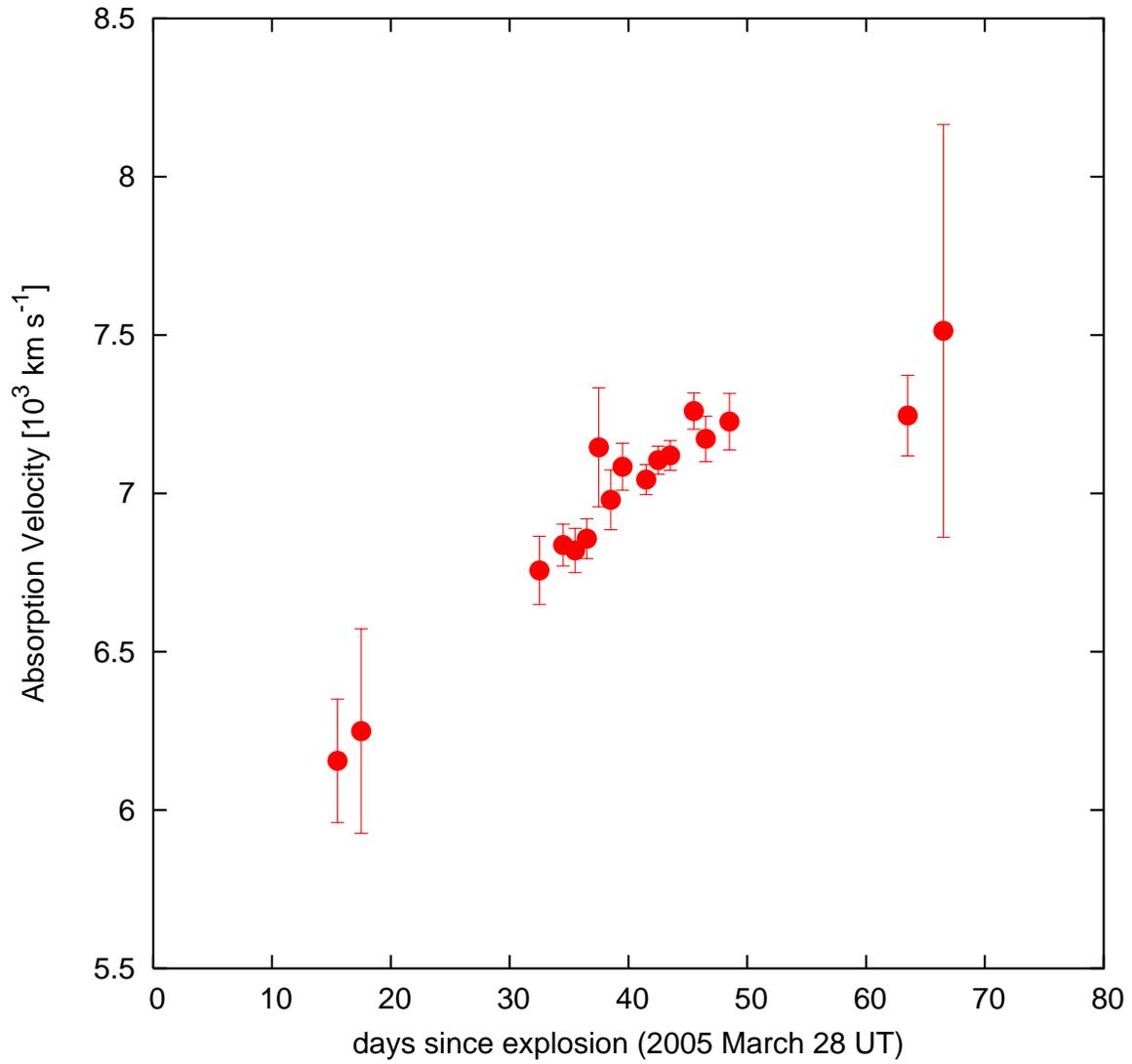} 
\caption{The time evolution of the He line
velocity derived from the absorption minimum of He~I $\lambda$~5876
\citep{mod05c}. 
\label{fig:Hevel}}
\end{figure*}

The He absorption lines first appeared at a velocity of $\sim 6,000$ \kms. The
velocity increased slowly with time, reaching $\sim 7,000$ \kms\
(Fig.~\ref{fig:Hevel}), an unusual behavior for SN Ib (see, e.g., Branch et al
2002). On the contrary, the velocities of the Fe lines declined with time.
H$\alpha$, and possibly H$\beta$ and H$\gamma$, were detected 
(\citealt{wan05,anu05}), as possibly in other SN Ib (Deng et al. 2000; Branch
et al. 2002). The weakness of these lines suggests a small H mass (see \S 3),
so we assumed that H did not affect the LC. Seeing traces of H suggests
that the He layer was fairly intact before the explosion, so we included it
in our LC computation. We assumed that He is mainly above $v\gsim 6,000$ \kms,
as suggested by the observed He line velocity.  The light curve for a model
with $\Mej=7\Msun$ and  $E_{51} = E/10^{51}{\rm ergs}=2.1$ is shown as a dashed
line in Figure~\ref{fig:LC}. 

The model WR star explosion of \cite{ens88} produced a double-peaked LC, but 
in that case the first peak was caused by recombination of He ionized by shock
heating, and is too narrow for the first peak of the light curve of SN~2005bf,
which has a width of $\sim 15$ days. The only possibility we found to reproduce
the double-peaked LC was to assume a distribution with most \Nifs\ at low
velocity, and a small amount at high velocity, as follows: ($X({\rm ^{56}Ni})$,
$M({\rm ^{56}Ni})/\Msun$) = (0.75, 0.18) at $v\lsim 1,600$ \kms, (0.015, 0.04)
at $v \sim 1,600 - 3,900$ \kms, (0.025, 0.07) at $v \sim 3,900 - 6,200$ \kms,
and no \Nifs\ at  $v \gsim 6,200$ \kms. Here $X$ denotes the mass fraction.  The
total \Nifs\ mass is $M({\rm ^{56}Ni})=0.29\Msun$. The first peak of the LC is
powered by a small amount of high-velocity \Nifs\ at $v \gsim 3,900$ \kms. The
second peak is powered by \Nifs\ in the low velocity central region. The outer
extent of the outer ($v\sim 6,200$ \kms) and inner ($v\sim 1,600$ \kms) \Nifs\
are well determined from the time of the first and second peak, but the inner
distribution of the outer \Nifs\ is not well constrained, since it has little
effect on the light curve.

The large amount of \Nifs\ ($\sim 0.29\Msun$) would ordinarily lead to a bright
\Cofs\ tail. The dashed line in Figure~\ref{fig:LC} shows a LC computed with a
$\gamma$-ray opacity $\kappa_\gamma=0.027$ ${\rm cm^2 g^{-1}}$ (Shigeyama \&
Nomoto 1990).  However, the observed LC continued to decline rapidly,
suggesting that $\gamma$-rays escape more easily from SN 2005bf than in the
typical case. We used a reduced $\gamma$-ray opacity to simulate this
situation.

Using $\kappa_\gamma = 0.001 {\rm cm^2 g^{-1}}$ at $v<5,400$\,\kms\ (but
keeping $\kappa_\gamma=0.027 {\rm cm^2 g^{-1}}$ at $v>5,400$\,\kms), we
obtained a set of models with ($\Mej/\Msun$, $E_{51}$) = (5, 0.6), (6, 1.0),
(7, 1.3), (8, 1.7), (9, 2.3), (10, 2.8), and (11, 3.3) that reproduce the light
curve. The models with $M_{\rm ej}= 6-7 \Msun$ best fit the spectra (\S 3). The
solid line in Figure~\ref{fig:LC} shows the LC with $\Mej=7\Msun$, $E_{51} =
1.3$, $M({\rm ^{56}Ni})=0.32\Msun$, and the following  \Nifs\ distribution:
($X({\rm ^{56}Ni})$, $M({\rm ^{56}Ni})/\Msun$) = (0.44, 0.22) at $v\lsim
1,600$\,\kms, (0.013, 0.04) at $v \sim 1,600 - 3,900$\,\kms, (0.03, 0.06) at $v
\sim 3,900 - 5,400$\,\kms, and no \Nifs\ at $v \gsim 5,400$\,\kms.

Enhanced $\gamma$-ray escape could result from the presence of low-density
``holes''.  Such a structure may be produced by jets and Rayleigh-Taylor
instabilities possibly caused by magnetar activity (see \S4). While
$\gamma$-rays could escape much more easily from the low-density holes, those
$\gamma$-rays that are trapped in the high-density region are converted to
optical photons. Thus optical photons are selectively produced in high  density
regions, where the optical opacity can be expected to be normal. 

Because of the smaller average $\kappa_\gamma$ in the \Nifs\ region and the
enhanced escape of $\gamma$-rays with respect to a model with normal
$\kappa_\gamma$,  more \Nifs\ is necessary to power the LC, but lower velocity
\Nifs\ (i.e., $v \lsim 5,400$ \kms) can reproduce the first peak. In other
words, \Nifs\ is mixed out but does not reach the He layer at $v \gsim 6,000$
\kms. Such a double-peaked distribution of \Nifs\ might be produced by jets
that did not reach the He layer.

\section{SPECTROSCOPIC MODELS}
\label{sec:spectra}

We computed synthetic spectra from the explosion models described above. The
comparison with the observed spectra is necessary to distinguish among the
models. We used the Monte-Carlo spectrum synthesis code described in Mazzali \&
Lucy (1993), Lucy (1999), and Mazzali (2000). Since the code does not take into
account non-thermal processes, which are essential to populate \ion{He}{1} levels
\citep{luc91}, we introduced a ``non-thermal factor'' (Harkness et al. 1987) to
reproduce the He lines. This parameterized factor, denoted as $f$, represents
the degree of departure from the level populations computed using a modified
nebular approximation, and is used to multiply all \ion{He}{1} line opacities. 

\begin{figure*}
\epsscale{1.8}
\plotone{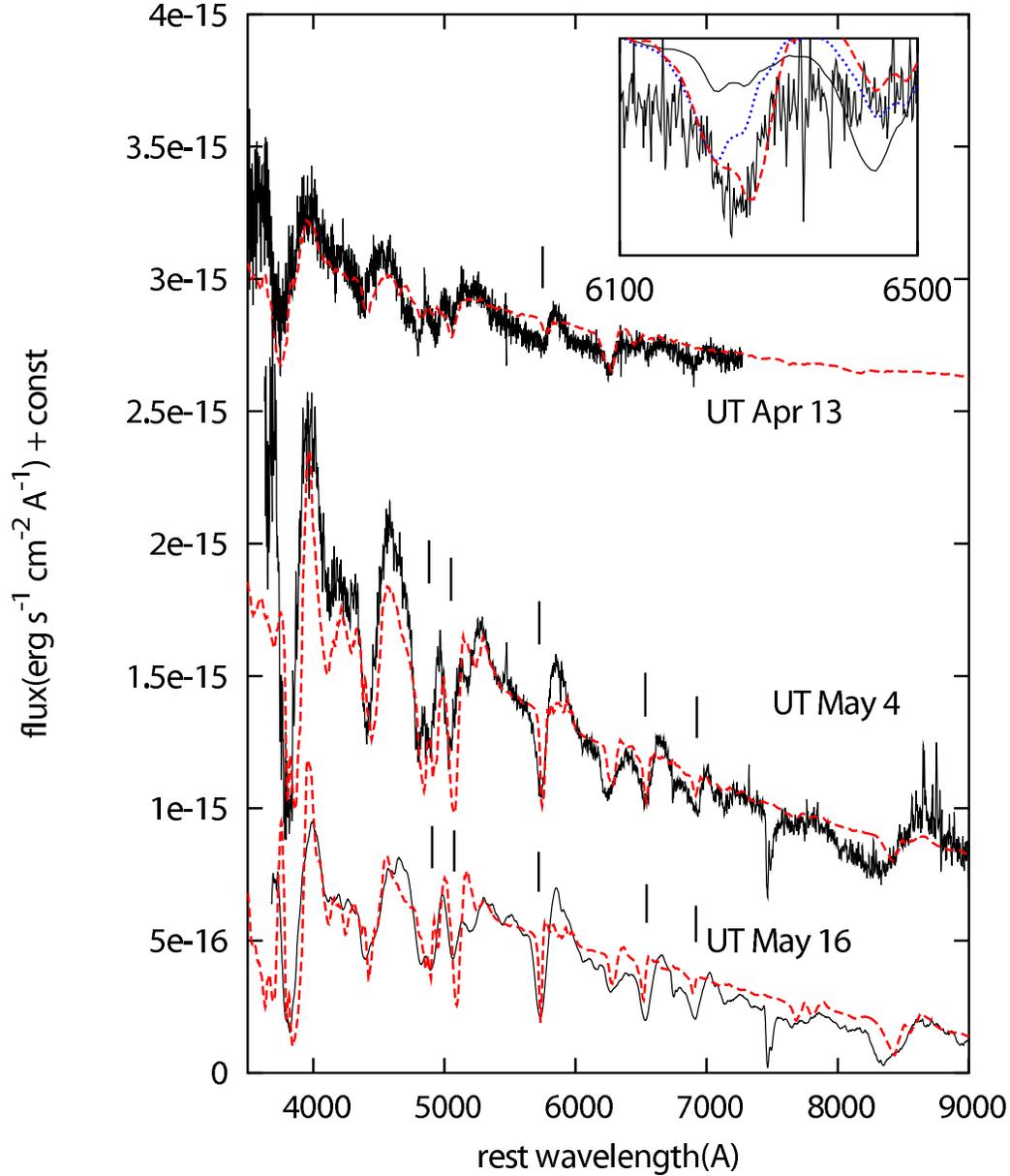} 
\caption{Spectra of SN~2005bf (thick lines: 
2005 April 13 - FLWO, Modjaz et al. 2005c; 
May 4 - HCT, Anupama et al. 2005;
May 16 - Subaru Telescope, Kawabata et al. 2005) compared to the
synthetic spectra (dashed lines) computed with the model ($\Mej/\Msun$,
$E_{51}$) = (7, 1.3). The position of He lines is shown by tick marks. 
The absorptions near 4900 and 5100\AA\ are blended with \ion{Fe}{2} lines. 
The inset shows the absorption near 6300\AA\ in the April 13th spectrum. 
The model with H at $v \gsim\ 13,000$ \kms\ (dashed line) provides the best 
fit. Thin and dotted lines show models with H in the whole ejecta and no H, 
respectively. \label{fig:spectra}} 
\end{figure*}

Our synthetic spectra are compared to the observed ones in Figure
\ref{fig:spectra}. The bolometric luminosities used to fit the spectra agree
with the observed ones within 10\%. The model with ($\Mej/\Msun$, $E_{51}$) =
(7, 1.3) provides satisfactory fits for all spectra. The photospheric
velocities this model predicts agree with those used to fit the spectra within
20\%. 

At the time of the first peak (UT April 13, 16 days after explosion; Modjaz et
al. 2005c) SN~2005bf exhibited SN Ic features, but actually both He and H can
be distinguished.  The feature near 5700\AA\ is probably \ion{He}{1} 5876\AA. 
We obtained a good fit for this spectrum with a bolometric luminosity $L = 2.0
\times 10^{42}$ \ergs\ and a photospheric velocity $v_{\rm ph}=6,200$\,\kms.
The non-thermal factor was set to 1. The model with $\Mej=7\Msun$ fits better
than that with $\Mej=6\Msun$. The feature at 6300\AA\ is reproduced as a blend
of H$\alpha$ and \ion{Si}{2} 6355\AA\ (Fig.~\ref{fig:spectra}, inset). The core
of the line cannot be due to \ion{Si}{2} because this would require a Si
velocity of $\sim 5,000$\,\kms, which is smaller than the photospheric
velocity.  It can, however, be reproduced as H$\alpha$ if $\sim 0.02\Msun$ of
H is present above $v \gsim 13,000$ \kms.  

Near maximum brightness (UT May 4, 37 days after explosion) He lines become
conspicuous. The model has $L = 5.0 \times 10^{42}$ \ergs, $v_{\rm ph}=4,600$
\kms, and $f=2.0 \times 10^{3}$  (only at $v \gsim 6,500 $ \kms).  Most
features, including \ion{Ca}{2}, \ion{Fe}{2} and \ion{Mg}{2} are also well
reproduced. The line near 6300\AA\ is reproduced as \ion{Si}{2} (6355\AA) at
this epoch: the photospheric velocity is low enough for Si to be above the
photosphere, while the H layer is too far above the photosphere to produce an
absorption. For a spectrum on UT May 16th (49 days, Kawabata et al. 2005),  we
also obtain a reasonable fit with parameters $L = 3.8 \times 10^{42}$ \ergs,
$v_{\rm ph} = 3,800$ \kms, and $f=2 \times10^6$ (at $v \gsim 7,200 $ \kms). 

These values of the ``non-thermal factor'' compare well with the results of
detailed calculations (Lucy 1991), although the value at the last epoch appears
too large. To mimic the increasing velocity of \ion{He}{1} $\lambda$ 5876, we
introduced the non-thermal factor at increasingly high velocity at more
advanced epochs. Enhanced $\gamma$-ray escape from the \Nifs-dominated region
can yield rapidly increasing $\gamma$-ray deposition in the He layer and higher
He excitation, as confirmed by the fact that the $\gamma$-ray deposition rates
derived from the LC calculations increase gradually over time at 7,200 \kms.
The ratios between the deposition at 7,200 \kms\ 49 days and 16 days after the
explosion is 14.5 (for comparison it is 4.6 at 6,200 \kms). As the optical
depth of the He lines increases, the region where they become optically thick
(and where the non-thermal factor is applied) moves to higher velocity.

\section{CONCLUSIONS \& DISCUSSION}
\label{sec:discuss} 

We have studied the properties of SN 2005bf by modelling the light curve and
the spectra. Our best fit model has $\Mej \sim 7\Msun$ and $E_{51} \sim 1.3$. 
The ejecta consist of \Nifs\ ($\sim 0.32 \Msun$), He ($\sim 0.4\Msun$),
intermediate mass elements (mainly O, Si, S), and a small amount of H ($\sim
0.02\Msun$). Thus the progenitor had lost almost all its H envelope, but
retained most of the He-rich layer, as in WN stars. The double-peaked light
curve is reproduced by a double-peaked \Nifs\ distribution: such a distribution
may be caused by jets that did not reach the He layer. Strong He lines were not
seen in the earliest spectra because radioactive $^{56}$Ni was too far from the
He layer to excite He atoms. Increasing $\gamma$-ray deposition in the He layer
due to enhanced $\gamma$-ray escape from the $^{56}$Ni-dominated region may
explain both the delayed strengthening and the increasing velocity of the He
lines.

The He core mass at explosion was $M_{\rm He}=\Mej+M_{\rm cut}\sim
7.5-8.5\Msun$. The progenitor was probably a WN star of main-sequence mass 
$\Mms \sim25-30\Msun$ \citep{nom88,ume05}.  The formation of a WN star from a
star of only $\sim 25\Msun$ suggests that rotation may have been important
\citep{hir05}, although not sufficient to make SN 2005bf a hypernova.

Our model requires the production of an unusually large amount of \Nifs\ but a
normal explosion energy. To examine whether this is possible, we performed
hydrodynamical simulations of explosive nucleosynthesis for the model with
$\Mms=25\Msun$ and $E_{51}=1.3$ as in \cite{ume05}.  In order to produce $\sim
0.32\Msun$ \Nifs, the mass cut that separates the ejecta and the compact
remnant should be as deep as $M_{\rm cut}\sim 1.4\Msun$.  In the spherical
model no fallback occurs.  This suggests that the remnant was a neutron star
rather than a black-hole.  The mass range $\Mms \sim 25\Msun$ is near the
transition from neutron star (SN 2005bf) to black hole formation (SN 2002ap;
Mazzali et al. 2002), the exact boundary depending on rotation and mass loss.

Interestingly, the progenitor of the Cassiopeia A (Cas A) SN remnant was also
probably a WN star, and its nucleosynthetic features are consistent with a
$\sim 25\Msun$ star (e.g., \citealt{fes05} and references therein).  The
compact remnant of Cas A could be a magnetar (a neutron star with strong
magnetic fields, Krause et al. 2005). The Cas A remnant is extremely clumpy,
with many knots.  The fastest Fe-rich 'jet' moves with $v \sim 6,000$ km
s$^{-1}$ and is not mixed with N-He or O-rich layers \citep{fes05}.  These
properties resemble our model for SN 2005bf.  We might speculate that SN 2005bf
was similar to the explosion of Cas A and that magnetorotational effects at
collapse and later magnetar activity produced jets that created extremely
clumpy ejecta in the $^{56}$Ni-rich layer.  These jets were not energetic
enough to reach the He layer.

\begin{figure*}
\epsscale{2.}
\plotone{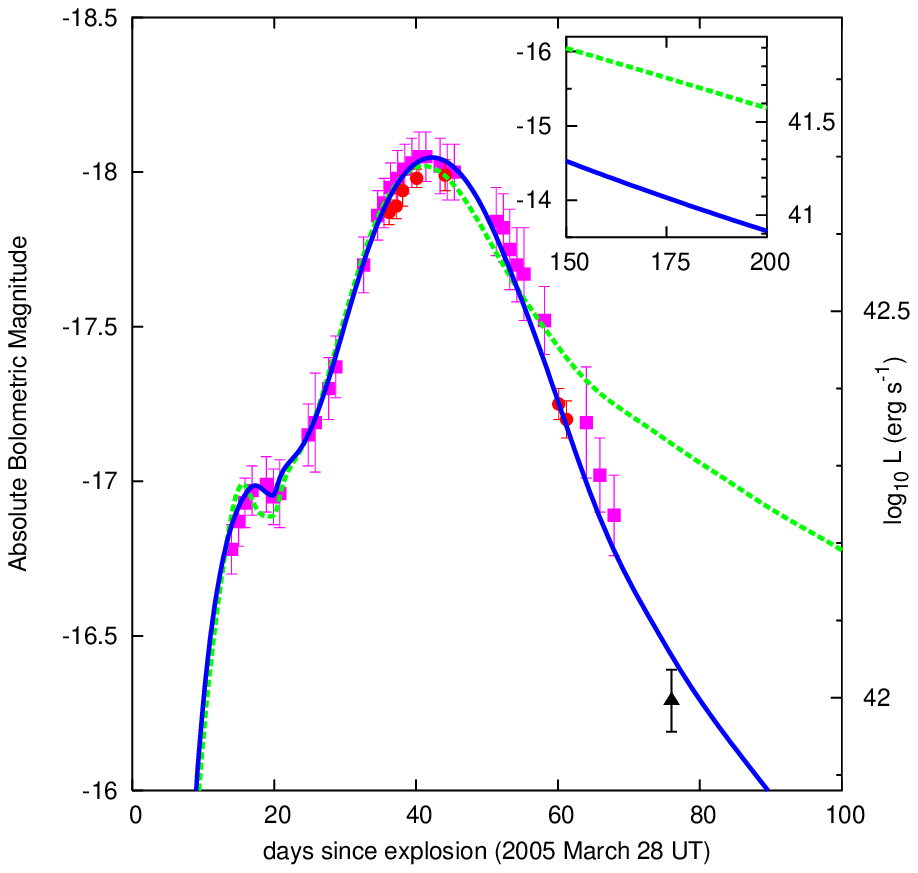} 
\caption{The bolometric light curve constructed from FLWO (filled squares; 
\citealt{mod05c}), HCT (filled circles; \citealt{anu05}), 
and MAGNUM (filled triangle; \citealt{ina05}) photometry. 
The contribution of near-IR light is between $\sim$ 20\% of the total at 
early phases to $\sim$ 50\% at late phases.
Synthetic light curves are shown for normal (dashed) and reduced (solid) 
$\gamma$-ray opacities (see text). The inset shows the predicted LCs of each 
model when SN~2005bf emerges from behind the Sun in fall 2005. 
\label{fig:LC}}
\end{figure*}

Our best-fit model with enhanced $\gamma$-ray escape predicts that the apparent
bolometric magnitude of SN 2005bf as it emerges from behind the Sun in late
2005 will be $\sim 20.6$ (Fig.~\ref{fig:LC}, inset).  Without the enhanced
$\gamma$-ray escape, this value would be $\sim 19.0$  mag. SN~2005bf showed
significant polarized spectral features \citep{wan05,kaw05}. Polarization
observations (e.g., \citealt{kaw02}) could be extremely useful to reveal the
detailed distribution of the elements in the ejecta and the orientation of the
possible jet, as would nebular spectra (Maeda et al. 2002; Mazzali et al.
2005).

\acknowledgments

This work has been supported in part by the Grant-in-Aid for Scientific
Research (16540229, 17030005, 17033002) from the MEXT of Japan.
Research on supernovae at Harvard University has been supported by NSF
Grant AST-0205808.  The PAIRITEL is operated by the Smithsonian
Astrophysical Observatory and a grant from the Harvard University Milton
Fund.

\clearpage

\clearpage

\clearpage

\clearpage

\clearpage

\begin{thebibliography}{}

\bibitem[Anupama \etal(2005)]{anu05} Anupama, G. C., Sahu, D. K., Deng,
                             J., Nomoto, K., Tominaga, N., Tanaka, M.,
                             Mazzali, P. A., \& Prabhu, T. P. 2005,
				ApJ, 631, L125
				
\bibitem[Arnett(1982)]{arn82} Arnett, W. D. 1982, \apj, 253, 785

\bibitem[Branch \etal(2002)]{bra02} Branch, D., et al., 2002, \apj, 566,
			     1005

\bibitem[Deng \etal(2000)]{den00} Deng, J.S., Qiu, Y.L., Hu, J.Y., Hatano, K.,
			     \& Branch, D. 2000, \apj, 540, 452

\bibitem[Ensman \& Woosley(1988)]{ens88} Ensman, L.M., \& Woosley,
			     S.E. 1988, \apj, 333, 754

\bibitem[Falco \etal(1999)]{fal99} Falco, E.E., \etal\ 1999, \pasp, 111, 438

\bibitem[Fesen \etal(2005)]{fes05} Fesen, R.A., \etal\ 2005, \apj, in press
			     (astro-ph/0509067)

\bibitem[Folatelli \etal(2005)]{fol05} Folatelli, G., \etal\ 2005, \apj,
			     Submitted (astro-ph/0509731)

\bibitem[Hamuy \etal(2005)]{ham05}Hamuy, M., Contreras, C., Gonzalez, S., 
			Krzeminski, W. 2005, \iaucirc\ 8520

\bibitem[Harkness \etal(1987)]{har87}Harkness, R.P., \etal\ 1987, \apj, 317, 355

\bibitem[Hirschi, Meynet \& Maeder(2005)]{hir05} Hirschi, R., Meynet,
			     G., \& Maeder, A. 2005, \aap, 433, 1013

\bibitem[Inada \etal(2005)]{ina05} Inada, N., \etal\ 2005, in preparation

\bibitem[Iwamoto \etal(2000)]{iwa00} Iwamoto, K., \etal\ 2000, \apj, 534, 660

\bibitem[Kawabata \etal(2002)]{kaw02} Kawabata, K., \etal\ 2002, \apj,
			     580, L39

\bibitem[Kawabata \etal(2005)]{kaw05} Kawabata, K., \etal\ 2005, in preparation

\bibitem[Krause \etal(2005)]{kra05} Krause, O., et al.  2005, Science, 308, 1604

\bibitem[Lucy(1991)]{luc91} Lucy, L. B. 1991, \apj, 383, 308

\bibitem[Lucy(1999)]{luc99} Lucy, L. B. 1999, \aap, 345, 211


\bibitem[Maeda \etal(2002)]{mae02} Maeda, K., Nakamura, T., Nomoto, K.,
			     Mazzali, P.A., Patat, F., \& Hachisu,
			     I. 2002, \apj, 565, 405

\bibitem[Mazzali \& Lucy(1993)]{maz93} Mazzali, P. A. \& Lucy,
                      L. B. 1993, \aap, 279, 447

\bibitem[Mazzali (2000)]{maz00} Mazzali, P. A. 2000, \aap, 363, 705

\bibitem[Mazzali \etal(2001)]{maz01} Mazzali, P. A., Nomoto, K.,
   Cappellaro, E., Nakamura, T., Umeda, H., \& Iwamoto, K. 2001, \apj, 547, 988
				 
\bibitem[Mazzali \etal(2002)]{maz02} Mazzali, P. A., et al. 2002, \apj, 572, L61
 
\bibitem[Mazzali \etal(2005)]{maz05} Mazzali, P. A., et al. 2005,
			     Science, 308, 1284
 

\bibitem[Modjaz \etal(2005a)]{mod05a} Modjaz, M., Kirshner, R., 
               Challis, P., Matheson, T., \&  Landt, H., 2005a, \iaucirc\ 8509
		       
\bibitem[Modjaz \etal(2005b)]{mod05b} Modjaz, M., Kirshner, R., 
                       \& Challis, P.  2005b, \iaucirc\  8522

\bibitem[Modjaz \etal(2005c)]{mod05c} Modjaz, M., \etal\ 2005c, in preparation 

\bibitem[Monard(2005)]{mon05} Monard, L. A. G. 2005, \iaucirc\ 8507

\bibitem[Moore \& Li(2005)]{moo05} Moore, M. \& Li, W. 2005, \iaucirc\ 8507

\bibitem[Morrell \etal(2005)]{mor05} Morrell, N., Hamuy, M.,
                       Folatelli, G., Contreras, C. 2005 \iaucirc\ 8509 

\bibitem[Nakamura \etal(2001)]{nak01} Nakamura, T., Mazzali, P. A.,
			Nomoto, K., Iwamoto, K. 2001, \apj, 550, 991

\bibitem[Nomoto \& Hashimoto (1988)]{nom88} Nomoto, K., \& Hashimoto, M.
                             1988, \physrep, 163, 13

\bibitem[Nomoto \etal (1995)]{nom95} Nomoto, K., Iwamoto, K., \&
                                Suzuki, T. 1995 Phys. Rep., 256, 173

\bibitem[Nomoto \etal (2004)]{nom04} Nomoto, K., Maeda, K., 
                                   Mazzali, P.A., Umeda, H.,
                                   Deng, J., \& Iwamoto, K. 2004, 
                                 in Stellar Collapse, ed. C. L. Fryer
                                 (Kluwer: Dordrecht), 277 (astro-ph/0308136)
                              

\bibitem[Schlegel, Finkbeiner, \& Davis(1998)]{sch98} Schlegel, D. J.,
			     Finkbeiner, D. P., \& Davis, M. 1998 \apj,
			     500, 525

\bibitem[Shigeyama \& Nomoto (1990)]{shi90} Shigeyama, T. \& Nomoto,
			     K. 1990, \apj, 360, 242

\bibitem[Umeda \& Nomoto (2005)]{ume05} Umeda, H., \& Nomoto, K. 2005,
				\apj, 619, 427

\bibitem[Wang \& Baade(2005)]{wan05} Wang, L., \& Baade, D.  2005,
                              \iaucirc\ 8521 

\bibitem[Yoshii \etal(2003)]{yos03} Yoshii, Y. \etal\ 2003, \apj, 592, 467



\end{thebibliography}
\end{document}